# Training Opportunities for Intelligent Transport Systems and Cooperative Intelligent Transport Systems

**Charis Chalkiadakis[1], Panagiotis Iordanopoulos[1], Evangelos Mitsakis[1]**

[1][1]Centre for Research and Technology Hellas - Hellenic Institute of Transport (CERTH - HIT)
E-mail: charcal@certh.gr

**Abstract**

Intelligent Transport Systems (ITS) and Cooperative Intelligent Transport Systems (C-ITS) are of high significance, mainly due to the benefits they have in terms of operation of the transport network. Despite ITS and C-ITS importance in the operation of the transport network, there is a major knowledge gap regarding their development, way of operation and significance worldwide and especially among the responsible for their deployment public authorities. In order for such fragmentations to be tackled, an online training platform concerning the operation and impacts of ITS and C-ITS has been designed in the framework of the European Union's Horizon 2020 funded CAPITAL project. In order for the proper design of the CAPITAL Online Training Platform, two main approaches have been studied: capacity building and massive open online courses. The present study provides insight regarding the design and the context of the CAPITAL Online Training Platform.

## *1. Introduction*

Intelligent Transport Systems (ITS) and Cooperative Intelligent Transport Systems (C-ITS) are an ever-growing sector not only in European countries but worldwide. Developments regarding ITS and C-ITS are continuous and new innovative solutions are introduced on a constant basis. Proof of this is the importance the European Union has put on these fields, through the funding provided for research in innovative ITS and C-ITS technologies the previous and recent years. Despite the proven importance of such technologies, there is a major gap of knowledge concerning their development, way of operation and significance worldwide and especially among the responsible, for their deployment, public authorities. One of the factors which lead to this knowledge gap, is due to the lack of training regarding ITS and C-ITS technologies, both at cultivating researchers and public authorities and at educating end users. To this end, a training platform, with an extensive educational database regarding ITS and C-ITS, has been developed in the framework of CAPITAL project.

CAPITAL is a project funded by HORIZON 2020 Research and Innovation Program. The CAPITAL project aims at spreading knowledge, in the fields of ITS and C-ITS, by creating and providing decision makers and related stakeholders with a series of educational opportunities in the form of Massive Open Online Courses (MOOCs). The courses provided are namely:

- Introduction to ITS and C-ITS.
- Communication technologies for ITS and C-ITS including relevant standards.
- TMC and roadside technologies for ITS.
- ITS and C-ITS user services.
- Impact assessment of ITS and impacts of selected ITS and C-ITS systems.
- Financial incentives, business models and procurement models for C-ITS deployment.
- Cost-benefit analyses of ITS services.
- Guidance in deploying ITS and C-ITS.
- Data protection and privacy.

Through the aforementioned courses, all relevant stakeholders have access to timely and reliable facts on ITS and C-ITS, through the provision of comprehensive information in the form of lectures.

The present paper aims at providing a detailed description of the CAPITAL online training platform, that has been formed under the European Union's Horizon 2020 Research and Innovation Programme, and its capabilities to cover the knowledge gap on ITS and C-ITS. A thorough presentation of the platform's available material is being developed in such a way as to all aspects that compose these systems and all the required knowledge that should be held by anyone who is interested in deploying ITS and C-ITS be clearly defined.

Moreover, the paper examines the preliminary results of the performance appraisal of CAPITAL online training platform on users. For this preliminary assessment of users' perspective towards this platform, a quantitative analysis using questionnaires is being developed and the results are being processed in a constructive way so that valuable conclusions about the continuous improvement of the platform will be exported. The questionnaire investigates the level of user's knowledge before and after their training through the platform and asks for feedback on future development of the platform.

## *2. Theoretical background*

In order for the CAPITAL Online Training Platform to be designed and developed, an initial literature review had been conducted regarding both the main principles of the Massive Open Online Courses (MOOCs) and Capacity Building.

### *2.1 MOOCs*

A MOOC is an online course that is accessible via the web, 24 hours a day and 7 days a week. MOOCs also offer and support unlimited participation. MOOCs also make use of both traditional and modern teaching materials like videos, readings, projects, and assignments (Luaran, 2013), (Kaplan & Haenlein, 2016).

Many MOOCs use the type of video lectures; just like the common way of teaching, but with the use of technological solutions (Yousef et al., 2015). The common duration of a course is among 5 - 10 minutes (University of Alicante, 2013). However, a study conducted by edX found that students stop watching videos longer than 6 to 9 minutes (L. Holmes, 2013).

Because of massive enrolments in e-learning platforms, MOOCs require an instructional design that facilitates large-scale feedback and interaction. The two basic approaches (Rivard, 2013), (University of Alicante, 2013) are:

- Peer-review and group collaboration.
- Automated feedback through assessments.

*2.2 Classification of MOOCs*

There are various types of MOOCs available. The two distinct categories are xMOOCs and cMOOCs; these two categories of MOOCs have main theoretical differences (Kesim & Aktinpulluk, 2014).

Below, there is a short description of xMOOCs. This is due to the reason that the CAPITAL Online Training Platform is based on the main characteristics of xMOOCs.

xMOOCs are the most common form of MOOCs and they are used to describe the courses developed by various online training/ educational platforms like Coursera and edX. xMOOCs are characterized by a variety of certain design features. The most common among them (Bates, 2015) are:

- xMOOCs use specially designed platform software with the ability for registration of large participants' number.
- xMOOCs use the standard lecture mode, despite they are a form of online courses.
- A lecture may last at about 12-15 minutes
- The total course duration is approximately 5 weeks.
- Students complete an online test after every week and the feedback is provided immediately.
- Teaching material may be available to the participants.
- There are forums where participants can post questions related to the course.
- After the completion of the course, and if the participant has a score above a minimum-set limit, a Certification of Attendance is provided.

*2.3 Capacity Building and Capacity Development*

Another important aspect regarding the design and the development of the platform is the investigation regarding the proper capacity building strategies needed to be implemented for the CAPITAL Online Training Platform to reach its goal.

Capacity-building is a term widely used in relation to different organizations. A general definition states: capacity-building is "*planned development of (or increase in) knowledge, output rate, management, skills, and other capabilities of an organization through acquisition, incentives, technology, and/or training.*" (Business Dictionary, n.d.).

The term witnessed increased interest in the 1990s. The 1993 UNDP report on Rethinking Technical Cooperation – Reforms for Capacity Building in Africa as the first attempt to define both capacity (in general) and capacity development (UNDP, 1993).

The 1996 Organization for Economic Co-operation and Development (OECD) report (OECD, 1996) introduced the term "capacity development". The introduction of such terms and their further analysis were inspired by the changes occurred in the development policy, in the late 1990s and the early 2000s. Some significant changes were of the UN Millennium Development Goals in 2000 and the Paris Declaration on Aid Effectiveness in 2005 (United Nations General Assembly, 2000), (OECD, 2008a). The latter highlights that capacity development is one of the essential preconditions for aid effectiveness: "*The capacity to plan, manage, implement, and account for results of policies and programmes, is critical for achieving development objectives – from analysis and dialogue through implementation, monitoring and evaluation*". Capacity-building, therefore, is the "*responsibility of partner countries*", while donors play a supporting role (OECD, 2008a).

Despite the existence of the aforementioned references regarding capacity, capacity building and capacity development the OECD Development Assistance Committee (DAC) study of 2006 (OECD, 2008b) includes the most widely accepted and used definitions of the said terms. It is also worth mentioned that in the said study (OECD, 2008b) there is a clear demarcation between the terms capacity building and capacity development. In this study (OECD, 2008b) capacity development is described as the set of techniques and methodologies applied in order to strengthen the knowledge of a society towards a specific subject. On the other hand, capacity building is the application of certain techniques and methods in order for knowledge upon a certain subject to be gained. The latter implies that no knowledge upon the specific study exists, so it is well-stated by the OECD (OECD, 2008b) that the term "capacity development" expresses better the concept of capacity and the efforts needed for its increase.

## *3. The CAPITAL on-line training platform*

The CAPITAL Online Training Platform is an attempt of the CAPITAL project consortium to develop and offer a training tool on ITS and C-ITS in order to assist their effort on learning more about ITS and C-ITS deployment. The platform aims at firstly, raising awareness about the benefits of ITS and C-ITS and also forming and extending career paths of interested parties by improving their skills and expanding their knowledge on the technical, business and policy aspects of ITS and C-ITS implementation.

The structure of the platform is the outcome of a thorough investigation on the needs and requirements of the main target groups for training and there is a distinct focus on training of public authorities as they are the responsible entities for ITS and C-ITS design in cities. This investigation pointed out the necessity for the provision of different levels of training depending on the previous expertise and experience of the trainees. For this reason, three levels of training are available to address beginner, intermediate and advance knowledge level.

Moreover, concerning the chosen training material to be provided, the above-mentioned analysis of needs helped also on this, as potential trainees and stakeholders relevant to these systems determined the priority topics through several steps of data collection and by using different qualitative and quantitative methods such as questionnaire-based surveys. The specific training interests that had been exported were combined to form the learning material that would be available through the platform. Nine topic studies were developed, its one of them containing video-lectures dealing with a specific topic and forming the study module. For every topic study a partner was set in charge and he is responsible for the provision of the appropriate information so as the theme of the topic study was best covered. Accompanying the video-lecture, a course syllabus had been developed including the training material.

For every topic study there are some introductory information about its duration, the proposed effort that someone should put in, the institution which offers the course and the level of knowledge that is addressed by the specific course. Also, a short description of the theme of the course is provided, as well as a brief presentation of the trainer and the trainer's expertise on ITS and C-ITS. The courses language is English but CAPITAL project partners are considering the translation of the learning material and import of subtitles in the video-lectures so as to cater a wider audience.

*Table 1: Courses provided in the CAPITAL Online Training Platform*

| Course | Title | Length |
|---|---|---|
| Topic Study 1 | Introduction to ITS and C-ITS | 2 study weeks |
| Topic Study 2 | ITS and C-ITS User Services | 4 study weeks |
| Topic Study 3 | TMC and roadside technologies for ITS | 4 study weeks |
| Topic Study 4 | Communication technologies for ITS and C-ITS including relevant standards | 4 study weeks |
| Topic Study 5 | Impact assessment of ITS and impacts of selected ITS and C-ITS systems | 4 study weeks |
| Topic Study 6 | Financial incentives, business models and procurement models for C-ITS deployment | 4 study weeks |
| Topic Study 7 | Cost-benefit analyses of ITS services | 4 study weeks |
| Topic Study 8 | Guidance in deploying ITS and C-ITS | 4 study weeks |
| Topic Study 9 | Data protection and privacy | 4 study weeks |

The content of every topic study, as stated in Table 1, is presented below in more detail.

**Topic study 1**: Introduction to ITS and C-ITS

This topic study aims at providing the fundamentals on ITS and C-ITS and it caters beginners who have almost no prior knowledge on ITS and C-ITS. The study module presents clear definitions of ITS and C-ITS with an emphasis on their distinction. Also, ITS technologies and functions, as well as C-ITS services, are being discussed. After the course, the trainee has the ability to recognize ITS and C-ITS systems and understand the usefulness of transition from ITS to C-ITS.

**Topic study 2**: ITS and C-ITS user services

The objective of this topic study is to offer a wide knowledge of ITS and C-ITS services that are nowadays available to both professional and non-professional drivers. A range of selected services is briefly presented as well as their capabilities in assisting safe driving and avoidance of fatalities. Additionally, four ITS services that are developed to manage professional drivers and authorities' cooperation are described in one of the study modules and the topic study ends with guidelines on proper effort for communicating the beneficiary aspect of ITS and C-ITS to the general public.

**Topic study 3**: TMC and roadside technologies for ITS

The topic study at TMC and roadside technologies for ITS focus on demonstrating the application of ITS in controlling and management of traffic. After a short description of the main features of urban traffic management and requirements in transport data collection, the importance of integrated traffic management and dynamic contribution of road-users on traffic management is emphasized.

**Topic study 4**: Standards, architectures and communication technologies for ITS and C-ITS.

Communication technologies and other key technologies for ITS proper operation are discussed in detail on this topic study. Also, there is an analysis of information flows on C-ITS networks and provision of the basic standards and standardization organizations that form the architectures of communication for these systems. Finally, the trainee has the potential to learn about the utilization of the above-mentioned standards and the process for ensuring interoperability.

**Topic study 5**: Impact assessment of ITS and C-ITS systems.

This topic study is dealing with the available evaluation approaches and techniques for ITS and C-ITS impact assessment and also present paradigms of social-health and environmental impacts of selected ITS and C-ITS. Also, the trainer presents key considerations of the evaluation method and the ways of collecting data in order to be used for the impact assessment.

**Topic study 6**: Financial incentives and business and procurement models for C-ITS deployment

The material of this study module tries to cover the business aspect of C-ITS by offering the theoretical background of business models and implement this concept to the transport systems.

**Topic study 7**: Cost-benefit analyses of ITS services

The methodology of cost-benefit analysis implementation on C-ITS is the main topic of this study module. Trainees have the opportunity to learn the specifications of CBA analysis such as the purpose of deployment, calculation steps, indicators and restraints on its development. Also, the topic study includes a chapter about costs and benefits identification and refers the practical application of CBA.

**Topic study 8**: Guidance in deploying ITS and C-ITS

The objective of this topic study is to provide all relevant information in order to inform interested parties about the policies and strategies concerning ITS and C-ITS. Mainly the European policy framework is presented and analysis on its adaptation on local strategies as well as implementation criteria and guidelines for transferability of C-ITS solutions among cities.

**Topic study 9**: Information security, data protection and privacy

Data protection and privacy in C-ITS is a critical issue concerning the legal implementation of these systems. There are also some other features that are gathered to the platform and are compatible with the capabilities that are offered by corresponding online training platforms.

Trainees have the opportunity to discuss their concerns and recommendations about lectures, provide their feedback or to interact with each other to exchange knowledge or make queries on the provided learning material.

A pre-course and a post-course survey accompany each one of the nine courses. The pre-course survey tries to investigate the professional background, the knowledge background on ITS and C-ITS, the working effort that someone may put in the course and finally the reason for attending the specific course. The post-course survey is an effort of the consortium that developed the platform to gain useful information for evaluating the online training tool. It includes questions seeking the satisfaction of participants in the training courses and the structure of the study modules as well as to rate their post-course acquired knowledge in comparison to their knowledge in the specific topic before taking the course. Moreover, after every video-lecture, the trainees are required to take a quiz associated with the previously attended training, so they have the ability to self-assess the knowledge gained after attending the video-lecture.

Also, every trainee can monitor its personalized progress on the course and be kept informed about the sections that he has already attend and its practice scores in all quizzes.

## *4. Acceptability and Usefulness of the context of the platform*

The following figures aim at providing information regarding the actual result of the CAPITAL Online Training Platform in the participants of training events organized in the framework of the CAPITAL project.

Collected feedback has been analyzed by using the statistical method of descriptive statistics. Descriptive statistics is "*a summary statistical method that quantitatively describes (or summarizes) features of collected information*" (Mann, 1995). The methodological framework used for the analysis of the collected feedback is depicted in the figure below.

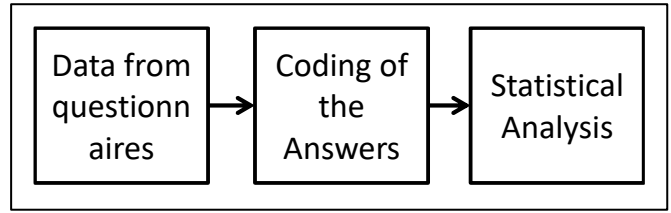

*Figure 1: Framework for the analysis of collected feedback*

The below figures aim at identifying the extent up to which the provided material at the training events help the participants at broadening their knowledge in the fields of ITS and C-ITS. The participants were categorized in 5 knowledge levels, according to their knowledge on ITS and C-ITS; Level 1 indicates that the participant is considered a "beginner" regarding ITS and C-ITS whereas Level 5 indicates that the participant is an ITS and/ or C-ITS expert.

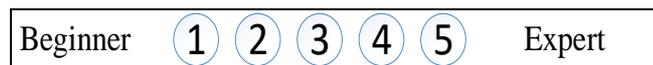

*Figure 2: ITS and/ or C-ITS stakeholders' levels of knowledge*

Before the training events, the participants were distributed in all 5 levels of knowledge. Interestingly, after the events, it is obvious that the knowledge of the participants is greater. Regarding ITS, the following statistics resulted after the analysis of the collected feedback.

*Table 2: ITS knowledge levels BEFORE and AFTER the training events*

| Knowledge level | Before the training | After the training |
|:---:|:---:|:---:|
| Level 1 | 7 | 0 |
| Level 2 | 4 | 5 |
| Level 3 | 21 | 6 |
| Level 4 | 5 | 17 |
| Level 5 | 2 | 11 |

The following figure (Figure 3) illustrates the information provided in Table 2.

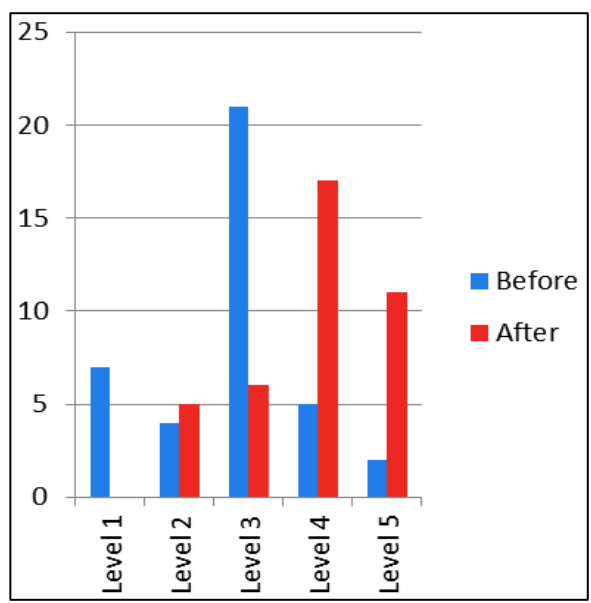

*Figure 3: Knowledge regarding ITS BEFORE and AFTER the CAPITAL training events*

The same changes in the distribution of the participants' knowledge level are observed regarding broadening their knowledge towards C-ITS.

In the following table (Table 3), as well as in the following figure (Figure 4), the redeployment of the participants among the 5 different knowledge levels is depicted.

*Table 3: C-ITS knowledge levels BEFORE and AFTER the training events*

| Knowledge level | Before the training | After the training |
|---|---|---|
| Level 1 | 9 | 0 |
| Level 2 | 2 | 6 |
| Level 3 | 22 | 6 |
| Level 4 | 4 | 15 |
| Level 5 | 2 | 12 |

The following figure (Figure 4) illustrates the information provided in Table 3 above.

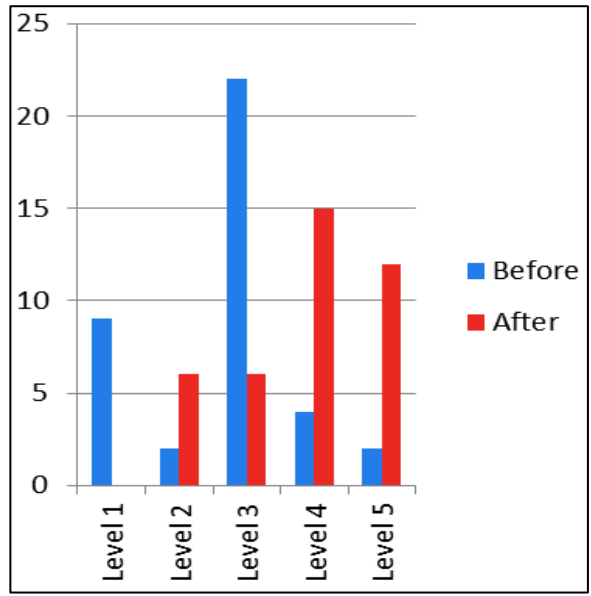

*Figure 4: Knowledge regarding C-ITS BEFORE and AFTER the CAPITAL training events*

It is obvious that after the training events took place, no participants considered themselves as "Level 1" ITS or C-ITS stakeholders, as well as there is an increase in the number of "Level 4" and "Level 5" participants; the significance of the provided material is therefore high towards broadening the knowledge of the participants in the ITS and C-ITS sector.

The usefulness of the CAPITAL Online Training Platform is also examined. The participants of the training events, after they interacted with the CAPITAL Online Training Platform during the training event, asked to state if they found the platform useful or not. The majority of the respondents stated that the CAPITAL Online Training Platform is useful.

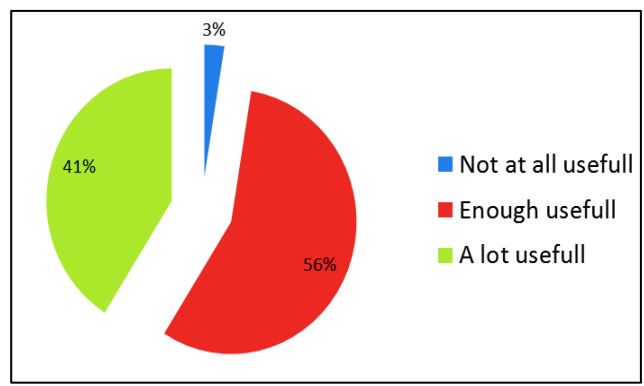

*Figure 5: Usefulness of the CAPITAL Online Training Platform*

The tables and figures presented above evince both the importance and the results CAPITAL Online Training Platform has on the participants of the training events. The knowledge upon ITS and C-ITS was broadened after the end of the events and almost all of the participants found the CAPITAL Online Training Platform and its contents useful.

## *5. Conclusions*

The significance of ITS and C-ITS technologies is major, not only in the EU level but, worldwide. Despite their significance though, there still exist a lack of information regarding ITS, C-ITS and their applications both in academic level and in the level of stakeholders.

Taking into account the above statement, it is therefore of high importance the existence of a database, or even better the existence of an online platform which is structured by considering both the main structure of MOOC platforms as well as the proper structure in order for someone to be educated and/ or trained regarding ITS and C-ITS.

The CAPITAL Online Training Platform integrates both of the aforementioned factors. It is structured by combining the main principles for designing an xMOOC online course and the main ideas and theories for educating and/ or training someone, despite their knowledge level, for ITS and C-ITS technologies.

From the conducted survey among the participants of the CAPITAL training events, it is obvious that the provision of such material to ITS and C-ITS stakeholders is beneficial towards enriching such stakeholders' knowledge in the fields of ITS and C-ITS.

## *Acknowledgements*

The current study has been executed in the framework of the CAPITAL project. The CAPITAL project has received funding from the European Union's Horizon 2020 Research and Innovation Programme under Grant Agreement number No 724106.

## *4. References-Bibliography*